\documentstyle[prl,twocolumn,aps]{revtex}
\begin{document}
\title {Adlayer core-level shifts of admetal monolayers on
transition metal substrates and their relation to the
surface chemical reactivity}
\author{D. Hennig}
\address{
Humboldt-Universit\"at zu Berlin, Institut f\"ur Physik, Unter der Linden 6,
D-10$\,$099 Berlin, Germany}
\author{M.V. Ganduglia-Pirovano
and M. Scheffler}
\address{Fritz-Haber-Institut der Max-Planck-Gesellschaft, Faradayweg
4-6, D-14$\,$195 Berlin-Dahlem, Germany}
\date{submitted November 6, 1995}
\maketitle

\begin{abstract}
Using density-functional-theory we study the electronic and structural
properties of a monolayer of Cu on the fcc (100) and (111) surfaces of the
late $4d$ transition metals, as well as a monolayer of Pd on Mo bcc(110).
We calculate the ground states of these systems,
as well as the difference of the ionization energies of an adlayer
core electron and a core electron of the clean surface of the adlayer
metal.
The theoretical results are compared to available
experimental data and discussed in a simple physical picture; it is shown why
and how adlayer core-level binding energy shifts
can be used to deduce information on the
adlayer's chemical reactivity.
\end{abstract}

\narrowtext
Modern catalysts typically consist of more than one component,  because the
activity and selectivity required in a specific chemical process may be
conveniently tuned by changing the catalyst's material composition.
\cite{sinfelt}
While the requirements to a specific catalyst are usually quite obvious, a
simple tool to analyze the reactivity of a multi-component system, which could
help in the design of new catalysts, is still missing, even when only  a coarse
scale is asked for.
Recently, Rodriguez, Campbell and Goodman investigated the core-level
shifts of supported monolayers of Ni, Cu, and Pd by X-ray photoelectron
spectroscopy (XPS) \cite{goodman1,goodman2,goodman3} in order to analyze the
physical and electronic properties responsible for the particular catalytic
properties of bimetallic surfaces.

The authors stress that parallel trends exist between the shifts of
core-electron binding energies of surface atoms of the supported monolayers
and the chemisorption properties of these systems. Furthermore, they note a
correlation between changes in CO desorption temperatures and relative core
level shifts, and suggest that core-level measurements may be a powerful
tool to examine the surface chemical reactivity.

We define the {\em adlayer core-level shift} (ACLS) as the difference of the
ionization energies of an adlayer core level and a core level of a
clean elemental crystal surface. The latter consists of the same element
as that forming the adlayer.
Thus, using the system Cu on Pd(100) as an example, the ACLS is
\begin{eqnarray}
\Delta_{\rm ACLS} & =
          & \tilde{I}_{c}^{\rm Cu/Pd(100)} - \tilde{I}_{c}^{\rm Cu(100)}
\nonumber\\
&  = & E^{\rm Cu/Pd(100)}(n_{c}=1) -E^{\rm Cu/Pd(100)}(n_{c}=2) \nonumber\\
& &-E^{\rm Cu(100)}(n_{c}=1)
+E^{\rm Cu(100)}(n_{c}=2)  \quad , \label{eq1}
\end{eqnarray}
\noindent where $\tilde{I}_{c}$ are the core-level ionization energies with
respect to the Fermi energies, and $E$ are total energies of the
ground state (two electrons in the core level ($n_{c}=2$))
and the excited state (one electron in the core level ($n_{c}=1$) and
an additional electron at the Fermi level).
The upper indices identify the considered systems, namely the
adsorbed Cu monolayer on a Pd(100) substrate and the clean Cu(100) surface.
It should be noted that the ACLS refers to an energy difference between
{\em two surfaces} consisting of the same element, which contrasts with the
definition
of surface core-level shift (SCLS), which
is the energy difference of {\em surface and bulk} ionization energies.
Therefore,
it might appear plausible that screening effects experienced by the core hole
may drop out in the ACLS. The above definition refers to Cu on Pd(100), but
a transcription to other systems is obvious.

The analysis of Rodriguez {\em et al.} \cite{goodman3}
simply ignored final state screening
contributions. Furthermore, they explained their results in terms
of charge transfer between the surface and substrate atoms. For example, they
argue that at a Cu monolayer adsorbed on an early transition metal (less than
half occupied $d$ band), some electron transfer takes place from  the overlayer
to the substrate, which will shift the Cu core levels towards higher binding
energies (positive ACLS). Similarly, a Cu monolayer
adsorbed on a late transition metal would {\em accept} electrons from
the substrate. This would produce a negative ACLS and a stronger Cu-CO
bonding. \cite{goodman3}

With respect to SCLSs it is by now clear that the charge transfer
picture and the initial-state approximation do not hold
\cite{dieter3,dieter2,alden95}; however, whether this criticism
also applies to ACLSs is not obvious. Nevertheless,
there exist severe doubts that the analysis and interpretation of Rodriguez
{\em et al.} can be trusted, although the idea to
design a tool to investigate surface reactivities is most appealing.

In this work we present a detailed density-functional-theory study of the
trends in ACLSs for adsorbed metallic monolayers and show that
also in ACLSs (similar to SCLSs) the final state screening contributions can
be significant.
We find practically no charge transfer {\em between}
adsorbate and substrate. However, there is a noticeable internal
polarization of the adlayer atoms. The results
show that and why a
correlation between ACLSs and surface reactivities exists.

We consider a monolayer of Cu on the fcc(100) and (111)
surfaces of Ru, Rh, Pd, and Ag, as well as a monolayer of Pd on Mo bcc(110).
Using Slater's transition-state concept to evaluate total energy differences,
we obtain from Eq. (\ref{eq1})
\begin{eqnarray}
\Delta_{\rm ACLS} & \approx &
-[\epsilon_{c}^{\rm Cu/Pd(100)}(n_{c}=1.5)
-\epsilon_{\rm F}^{\rm Cu/Pd}] \nonumber\\
& & +[\epsilon_{c}^{\rm Cu(100)}(n_{c}=1.5)
-\epsilon_{\rm F}^{\rm Cu}]  \quad , \label{eq2}
\end{eqnarray}
\noindent where
$\epsilon_{c}^{\rm Cu/Pd(100)}$ and $\epsilon_{c}^{\rm Cu(100)}$
denote the
Kohn-Sham eigenvalues of a particular core state of an adlayer Cu atom and
a surface atom of clean Cu(100).
The Fermi energies of the Cu/Pd(100) and Cu(100) slabs are
$\epsilon_{\rm F}^{\rm Cu/Pd}$ and $\epsilon_{\rm F}^{\rm Cu}$,
respectively.
In the initial-state approximation, the ACLS
is given by Eq. (\ref{eq2}) with $n_c=2$.

The electron density, surface atomic structure, total energies,
core-electron eigenvalues, and ACLSs  are calculated
using the full-potential linear muffin-tin orbital (FP-LMTO) method
together with the local-density approximation of the
exchange-correlation functional. \cite{perdew81} Details of the method are
described elsewhere. \cite{methfessel92}
The surfaces are modeled by slabs which have a thickness of nine-layers for
the fcc(100) and bcc(110) systems,  and seven-layers for the  fcc(111) ones.
To describe the transition-state, self-consistent electronic structure
calculations are performed under the constraint of charge neutrality, which
implies that half a valence electron is added at the Fermi level. This
describes the situation of an electronically fully relaxed final state and
thus applies to systems which have a Fermi reservoir of electrons.
The impurity problem of the localized core-hole is treated by means of a
supercell, such that the atom containing the core  hole is surrounded by
non-ionized nearest neighbor atoms.

For a qualified theoretical analysis of ACLSs it is important
to use the correct ground state geometry. Thus,
we have determined the surface relaxation of the outermost layer by
minimizing the total energy, while keeping the interatomic distances of
deeper
layers as those obtained from bulk calculations.
This approximation is well justified because the change in the first
interlayer spacing of the clean surface or the distance between the
adlayer and the first substrate layer would be less than $\pm 2\%$ of
the ideal interlayer spacing if the two topmost layers are allowed to relax
simultaneously.
For example, we have obtained for the clean Cu(111)
surface  a 0.9\%  and a 0.6\% contraction of the first
and second interlayer spacings. This implies an increase of 1.5\% of
the first interlayer spacing if compared with the case where just the
surface layer was allowed to relax (see Table I).
Experimentally a 0.7\% contraction has been observed \cite{cu111}.
Similarly, for Cu/Ru fcc(111) a contraction
of 2.6\% of the adlayer separation and a 3.1\% contraction of the
outermost Ru-layer separation has been obtained.
The latter result implies an increase of
only 0.2\% of the adlayer separation if compared with the case in which
just the adlayer was allowed to relax (see Table I).
These numbers compared well with
earlier calculations which found 2.5\% and 3.5\% contractions for the
adlayer and first substrate layer of Cu/Ru hcp(0001), respectively.
\cite{Feibelman}.
The results for the topmost layer relaxation calculations are summarized in
Table \ref{tab1}. In particular for Cu/Pd and Cu/Ag the surface relaxation is
substantial, reflecting the fact that the covalent radius of Cu is
significantly smaller than that of Pd and Ag. The Ru substrate is considered
here in the fcc structure because we do not expect that the difference between
the fcc and hcp structures is important for the problems discussed in this
paper. Obviously, the theory for Ru fcc(111) is compared with experiments for
Ru hcp(0001).

Table \ref{tab2} lists the ACLSs for  Cu $2p$ electrons for a
Cu monolayer on the fcc(100) and (111) surfaces of Ru, Rh, Pd, and Ag
and for Pd $3d$ electrons for a Pd monolayer on Mo bcc(110). The agreement
between calculated ACLSs and the few experimental data is good. In
particular, it is better for the reported $3d$ level
measurements than for the $2p$ level ones, which in turn have a
substantially larger lifetime broadening.
The theoretical results show  that for a monolayer of Cu
supported on late transition metals, the Cu core electrons are less strongly
bound (relative to the Fermi level) than those of clean Cu.
The magnitude of the ACLS increases with the filling of the substrate $d$ band.
The influence of the substrate
is indeed significant: For Cu/Pd the Cu $2p$ electron is by 0.7 eV weaker
bound than that of clean Cu.
The results for Cu/Ag do not follow this trend, i.e., for these systems the
magnitudes of the ACLSs are smaller than those of Cu/Pd, Cu/Rh and Cu/Ru.
But the adlayer core levels are still at smaller binding
energies than the corresponding levels at the  clean Cu surface.

The trend of the calculated ACLSs is already present in the initial-state
approximation (column 4 in Table II), but the difference between
the total ACLS  and the initial-state contribution,
which is the screening contribution
(column 5 in Table II), is obviously not negligible: For Cu/Ag(100)
it amounts to 0.3 eV and reduces the initial state contribution by
60\%.

We will first discuss the origin of the trend of the core-electron
eigenvalues of the non-excited
systems in their electronic ground states, i.e., the
initial-state contribution to ACLSs.
The theoretical analysis shows that the trend of the initial-state eigenvalues
is due to a change in the surface potential caused by a rearrangement within
the surface density of states (DOS): Electrons are
transferred from the top of the $d$ band of the Cu adlayer
into $sp$ DOS. This increases the bond strength between surface
layer and substrate and still enables the system to maintain local charge
neutrality. The $sp$ electrons in turn
spill out into the vacuum to reduce their
kinetic energy, which implies an induced surface dipole moment which has the
negative end at the vacuum side.
The emptying of the upper Cu $3d$ DOS at the surface is
consistent with the shift of the surface $d$ DOS towards the Fermi
level.

Figure
\ref{d_DOS} displays the $d$ contribution of the surface DOS at a clean Cu
(100) surface and at the Cu monolayer adsorbed on the fcc(100) surface
of Rh, Pd, and Ag. The mentioned shift to lower binding energies is clearly
visible. The core electrons feel this reduced $d$ electron density and also
shift towards the Fermi level, i.e., to lower binding energies.
Figure \ref{fig1} displays the changes of the electron
charges contained in the different muffin tin spheres. We present the
charges in ``empty spheres'' just above the surface in the vacuum,
the differences of the charges contained in the spheres of the Cu layer and a
sphere in Cu bulk, and  differences of the charges contained in the spheres of
the second layer, i.e., the top layer of the substrate, and the same atom in
the bulk. The clean Cu surface is treated correspondingly, namely as a Cu
monolayer on a Cu substrate. In the top substrate layers (full circles in Fig.
\ref{fig1}), the  charges are very similar to those in the bulk:
For Cu/Cu and Cu/Rh the deviations from the bulk charges are practically zero,
for Cu/Ru there are slightly less electrons and for Cu/Pd and Cu/Ag there are
slightly more electrons at the top substrate layer.
Bigger changes are present in the surface layer: For all surface layers
the number of electrons is reduced and this reduction is nearly compensated by
the electrons which we find in the ``empty spheres'' just outside the surface.
These results confirm the above explanation, namely that the these
systems ought to be described in terms of a rearrangement within (or
polarization of) the adlayer DOS as opposed to a charge transfer between the
surface layer and the substrate.

In distinction to the Cu/X results (X = Ru, Rh, Pd and Ag) the
calculations for a Pd monolayer supported on Mo bcc(110) reveal that the Pd
$3d$ core electron is more strongly bound in Pd/Mo than in Pd fcc(100).
Thus, for
Pd/Mo we find a positive ACLS.
For Pd/Mo the surface $d$ band is broader than that of clean Pd
the bottom is dominated by Pd $d$ states and the top by the
tails of the $d$ states of the Mo substrate.
Local charge neutrality
requires now a rearrangement of the $d$ and $sp$ electrons in a way  that the
surface $d$ DOS shifts to higher binding energy.
Similarly, inspection of the surface $d$ DOS for the Cu/X systems
(see e.g. Fig. \ref{d_DOS}) show that the $d$ band width
for clean Cu is wider than for the Cu/X systems.
In this case, one argues that the
narrower overlayer $d$ DOS must shift to lower binding energies so that
local charge neutrality is maintained.

The ACLSs are sometimes substantially different to this initial-state
effect;
we note however, that in the studied examples the {\em sign} of the initial
state contribution and that of the total ACLS is always the same; in general
even this does not hold necessarily, but we expect that it will hold for most
systems. We like to add that the screening contributions to ACLSs are
particularly large for Cu because noble metals have a low density
of states at the Fermi level and thus the screening of a core hole
by intraatomic polarization is rather unefficient.
Thus, screening due to the surrounding atoms is important.

Both, the initial-state and the screening contributions to the ACLS probe the
importance of $d$ states at the Fermi level.
A negative initial-state
contribution means that the adlayer Cu (or Pd) atoms $d$ band
is shifted closer to the Fermi level
(see Fig. \ref{d_DOS}), and a negative
screening contribution means that screening of a core hole in the adlayer is
better than for the surface atoms of clean Cu (or Pd).
A positive screening contribution
means that screening of a surface core hole is worse for the Cu (or Pd)
adlayer atoms
than for the clean Cu (or Pd) surface.

It is well known, that the $d$ states at the Fermi level determine the
reactivity of surfaces (see for example
\cite{hoffmann,Fukui,morrel94,hammer95}):
When there are $d$ states at the Fermi level, strong chemical bonds can be
formed with adsorbates, \cite{hoffmann,morrel94} and
Hammer and Scheffler have recently shown, that a low DOS just below the
Fermi level gives rise to an energy barrier hindering the dissociation
of H$_2$, while a high DOS just below $\epsilon_{\rm F}$ lowers or
removes the energy barrier. \cite{hammer95}
Thus, the difference in $\rm
H_2$ dissociation barriers observed
on pure Ni and Cu surfaces, arises due to the difference in the
positions of the substrate $d$ levels relative to the Fermi level on the
two surfaces. \cite{hammer95}

The results presented here demonstrate that the shift of core-electron
eigenvalues
sensitively detect the shift of the valence $d$ band.
One important conclusion that can be drawn from this result is that
as the $d$ states at the Fermi level typically act as  ``frontier
orbitals'' \cite{Fukui} in surface chemical reactions \cite{morrel94}, ACLSs
may indeed
give important information on the changes of the surface reactivity induced by
adsorbed metallic layers. Our above analysis, for example, suggests that the
reactivity of a Cu monolayer on a rhodium (or Pd or Ag) substrate is higher
than that of a
clean Cu surface, and the reactivity of a Pd monolayer on Mo is lower than
that of a clean Pd surface. It would be interesting to see if these
predictions will be verified experimentally.

This work is part of the program of the Consortium on ``The Physical
Aspects of Surface Chemistry on Metals, Alloys, and Intermetallics".
M.V. Ganduglia-Pirovano would like to thank P. Fulde
for his hospitality at the Max Planck
Institute in Stuttgart, where part of this work has been carried out.
Part of this work was supported by Sonderforschungsbereich 290 of the
Deutsche Forschungsgemeinschaft.

After our work was submitted we learned that the Cu/Rh and Cu/Ru surfaces do
indeed show higher chemical activities
than clean Cu surfaces for the CO oxidation and
cyclohexane dehydrogenation reactions, respectively.
\cite{goodman4,goodman5} We are grateful to D.W. Goodman for this
information.

\begin{table}
\begin{tabular}{c c c}
 &\multicolumn{2}{c}{surface relaxation}\\
material&fcc(100)&fcc(111)\\\hline
clean Cu &$-2.0$&$-2.4$\\
Cu/Ru&$-4.8$&$-2.8$\\
Cu/Rh&$-7.1$&$-5.0$\\
Cu/Pd&$-12.7$&$-9.3$\\
Cu/Ag&$-16.6$&$-10.9$\\
Pd/Mo&\multicolumn{2}{c}{$-1.5$ [bcc(110)]}\\
\end{tabular}
\caption{Relaxation of the outermost layer
in percentage of the interlayer spacing in the substrate bulk.}
\label{tab1}
\end{table}

{\onecolumn
\begin{table}
\begin{tabular}{c c | l l l   c }
& & ACLS   & initial-state & screening & ACLS \\
material  & surface  & theory    & contribution  & contribution&
experiment\\\hline
Cu/Ru    &fcc(100)   &$-0.42 $ &$-0.16 $    &$-0.26 $   & \\
         &fcc(111)   &$-0.42 $ &$-0.30 $    &$-0.12 $
&$-0.13$ [hcp(0001)] \\
Cu/Rh    &fcc(100)   &$-0.59 $ &$-0.46 $    &$-0.13 $
&$-0.43$ \\
         &fcc(111)   &$-0.57 $ &$-0.53 $    &$-0.04 $   & \\
Cu/Pd    &fcc(100)   &$-0.69 $ &$-0.85 $    &+0.16      & \\
         &fcc(111)   &$-0.67 $ &$-0.84 $    &+0.17      & \\
Cu/Ag    &fcc(100)   &$-0.18 $ &$-0.46 $    &+0.28      & \\
         &fcc(111)   &$-0.34 $ &$-0.53 $    &+0.19      & \\
Pd/Mo    &bcc(110)   &$+0.90 $ &$+0.77 $    &$+0.13$
&$+0.90$ \\
\end{tabular}
\caption{Calculated adlayer core-level shifts (ACLSs) of Cu $2p$ and Pd $3d$
levels, their
initial-state and screening contributions, and experimental
data from Ref. {\protect{\cite{goodman1}}}. All values are in eV. For the
definition of ACLSs see  text and Eqs. (1), (2).
}
\label{tab2}
\end{table}}

\onecolumn
\vspace*{4.5cm}
\begin{picture}(3,10)
\put(.0,.0){\includegraphics{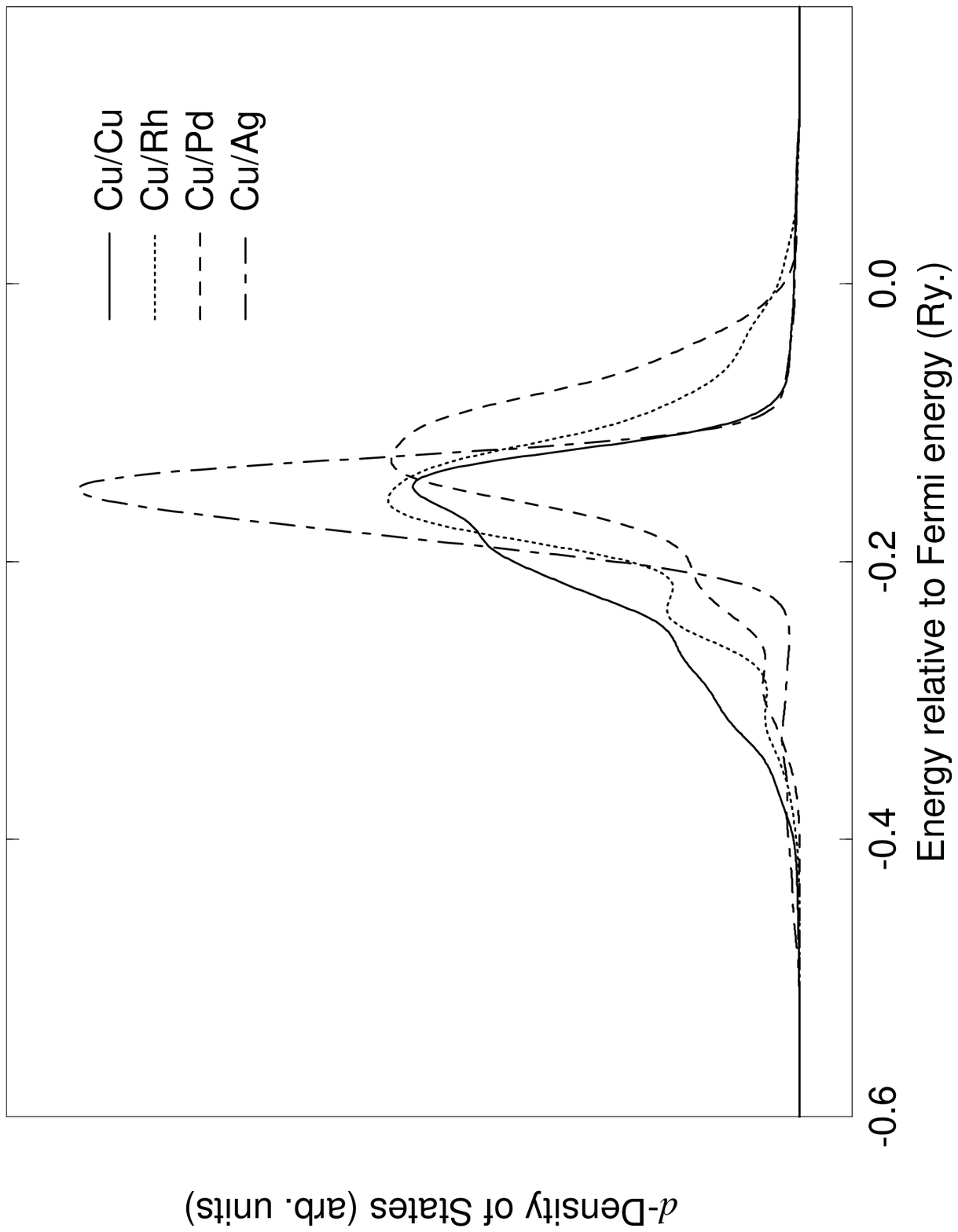}}
\end{picture}
\vspace*{-1.5cm}
\begin{figure}[h]
\caption{
Partial density of states, ($d$ contribution integrated over the muffin tin
sphere of the outermost atomic layer) for a clean surface of Cu fcc(100)
(solid line) and for a Cu monolayer on Rh fcc(100) (dotted line), on Pd
fcc(100)
(dashed line), and on Ag fcc(100) (dot-dashed line).
}\label{d_DOS}
\end{figure}

\vspace*{4.5cm}
\begin{picture}(3,10)
\put(.0,.0){\includegraphics{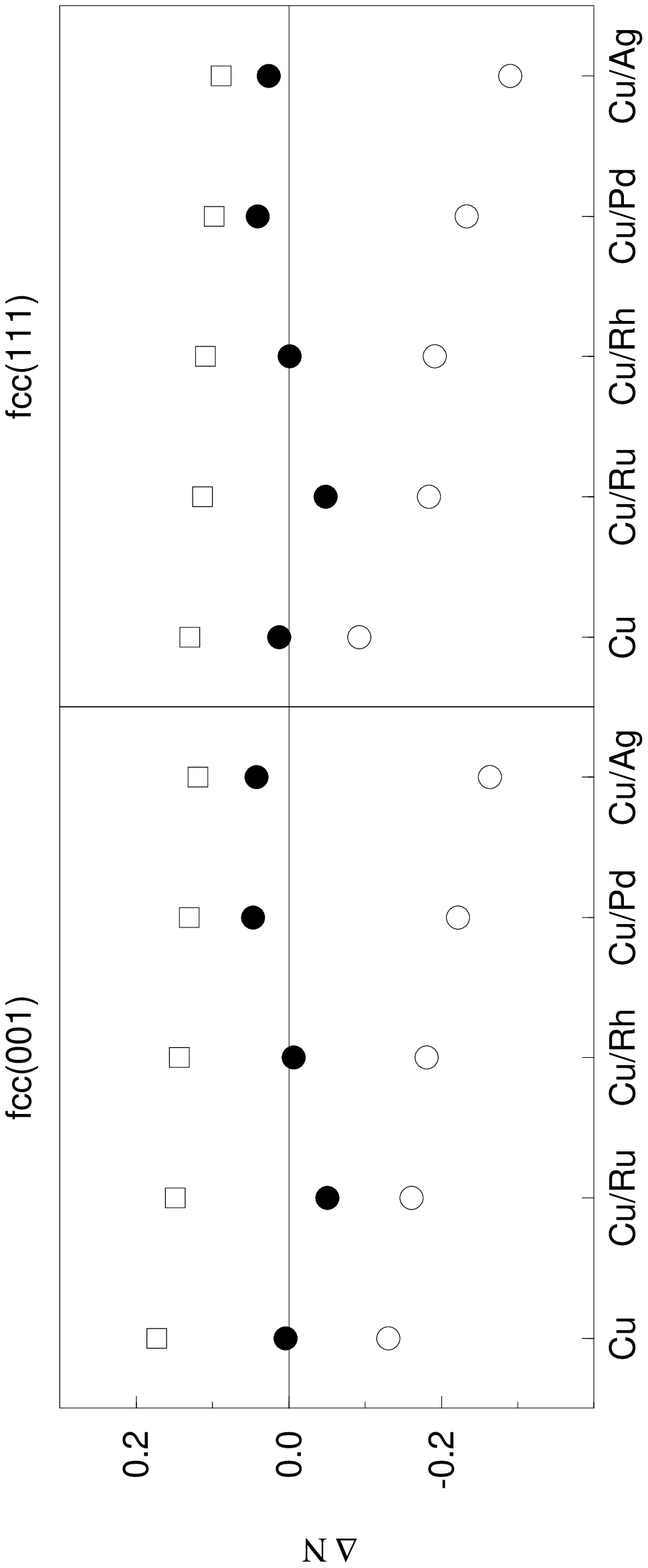}}
\end{picture}
\vspace*{-0.1cm}
\begin{figure}[h]
\caption{
Charge deficiency (negative $\Delta {\rm N}$) or charge increase (positive
$\Delta {\rm N}$) on individual atomic sites, compared to the charge in the
bulk
metals.
Full circles correspond to atoms in the
interface, i.e., in the layer touching the adlayer or the second
substrate layer for clean surfaces; open circles are for the adlayer or
surface layer. Open squares are for the layer representing the
vacuum-solid interface.
}\label{fig1}
\end{figure}

\end{document}